\documentclass{elsart}
\usepackage[dvips]{graphicx}

\begin{document}
\begin{frontmatter}
\title
{Microcanonical solution of the mean-field $\phi^4$ model:
comparison with time averages at finite size}
\author{Alessandro Campa\thanksref{mail1}}
\address{Complex Systems and Theoretical Physics Unit,\\
Health and Technology Department, Istituto Superiore di Sanit\`a,\\
and INFN Roma1, Gruppo Collegato Sanit\`a,\\
Viale Regina Elena 299, 00161 Roma, Italy}
\author{Stefano Ruffo\thanksref{mail2}}
\address{Dipartimento di Energetica ``Sergio Stecco'' and CSDC,\\
Universit\`a di Firenze, and INFN,\\
Via S. Marta 3, 50139 Firenze, Italy}

\thanks[mail1]{e-mail:campa@iss.infn.it}
\thanks[mail2]{e-mail:stefano.ruffo@unifi.it}

\begin{abstract}
We solve the mean-field $\phi^4$ model in an external magnetic field in the
microcanonical ensemble using two different methods. 
The first one is based on Rugh's microcanonical formalism and leads to express
macroscopic observables, such as temperature, specific heat, magnetization 
and susceptibility, as time averages of convenient functions of the phase-space. 
The approach is applicable for any finite number of particles $N$. 
The second method uses large deviation techniques and allows us to derive 
explicit expressions for microcanonical entropy and for macroscopic
observables in the $N \to\infty$ limit. Assuming ergodicity, we evaluate
time averages in molecular dynamics simulations and, using Rugh's approach,
we determine the value of macroscopic observables at finite $N$.
These averages are affected by a slow time evolution, often observed in 
systems with long-range interactions.
We then show how the finite $N$ time averages of macroscopic
observables converge to their corresponding $N\to\infty$ values
as $N$ is increased. 
As expected, finite size effects scale as $N^{-1}$. 
\end{abstract}

\begin{keyword}
Microcanonical ensemble; Mean-field models; Finite size effects

PACS: 05.20.-y, 05.20.Gg, 05.70.Fh
\end{keyword}

\end{frontmatter}

\section{Introduction}
The microcanonical ensemble can be considered, in several respects, 
as the basic ensemble for the description of the statistical behavior 
of physical systems. However, the canonical ensemble is often more 
amenable to exact or approximate analytical treatments 
and, therefore, several calculations are usually performed in its framework. 
On the other hand, there are some reasons of principle, and also of practical 
importance, for which the microcanonical ensemble must be given a prominent 
role~\cite{gross}. Indeed, there is an increasing evidence that ensembles can be
nonequivalent for systems with long-range interactions~\cite{LesHouches} or for
finite (small) systems, when surface effects are relevant~\cite{gross,chomaz}. 
In the mathematical physics literature, this question has been first discussed
in the pioneering paper by Hertel and Thirring~\cite{Thirring} (a recent publication
where several references to these results can be found is Ref.~\cite{Kiessling}).
An alternative approach, in which a parallel study of the distributions
of microstates and macrostates is performed, has been proposed in Ref.~\cite{ellis}. 
This approach is based on large deviation techniques~\cite{ellislibro}.

On the other hand, it has been shown that inequivalence between microcanonical and 
canonical ensembles can be associated to the presence of first order phase transitions 
in the canonical ensemble~\cite{barre}, and that there are regions in the phase diagram 
of the system that can be reached only within the framework of the microcanonical 
ensemble~\cite{touch,ellisb,Costeniuc}.
This means that in such cases, preparing the system at either fixed energy or temperature 
determines different values of the macroscopic observables. It has been 
correctly pointed out that this implies the impossibility of defining a unique equation of state for the system~\cite{chomaz}.
Moreover, in systems with long-range forces or in small systems, when the interaction between
macroscopic  subsystems cannot be reduced to a surface effect, the very concept of a heat bath can be  seriuously questioned~\cite{lynd2}.

The physical consequences of ensemble inequivalence are
of paramount importance. Perhaps, the most striking example is the existence of 
negative specific heat in the microcanonical ensemble for
self-gravitating systems~\cite{lynd1} . This reflects the real physical situation, 
and, in the canonical ensemble, negative specific heat is simply impossible.
Experimental observations of negative specific heat for atomic and molecular clusters
have also been recently reported~\cite{Schmidt,Gobet}.

The above arguments, which are all consequences of ensemble inequivalence, justify the necessity 
to perform, at least in some cases, microcanonical ensemble calculations. 
The operative problem then arises on how this can be actually done. 
This question can be faced both in the thermodynamic limit and at a finite number 
$N$ of degrees of freedom. The study of finite $N$ is relevant in two respects: 
$i)$ in Molecular Dynamics (MD) simulations one would like to estimate the approach 
to the $N\to\infty$ limit in numerical experiments, 
$ii)$ finite (small) systems, as just noted, can be interesting by themselves.

In this paper we discuss two different approaches to the study of a physical system
within the framework of the microcanonical ensemble. We will consider the
mean-field $\phi^4$ model in an external magnetic field.
We will here restrict to the description of the solution methods and to the implementation
of the expressions we obtain for macroscopic observables
in MD simulations. We will not discuss issues directly related to
ensemble inequivalence, which will be the subject of a forthcoming publication~\cite{CRT}.

In Section~\ref{twow} we review two different approaches to the 
calculation of microcanonical averages. The first one is based on the 
microcanonical formalism introduced by Rugh~\cite{rugh1,rugh2}. The second~\cite{ellisa} 
relies on large deviations techniques, which are particularly suited 
for mean-field like systems.
In Section~\ref{model} we present and briefly discuss the $\phi^4$ model.
In Section~\ref{implem} we show how the two approaches can be applied to the model.
In Subsection~\ref{averages} we derive the finite $N$ formulas for some observables in
Rugh's microcanonical formalism, while in Subsection~\ref{solution} large deviations techniques
are used to compute the entropy of the system, and, from it, the expressions of the thermodynamic
observables.
We then show that the large $N$ limit of the microcanonical formalism gives the same results
as the large deviation approach. In Section~\ref{results} we present the results of MD  simulations of our model, with the aim of illustrating the
approach to the $N \to \infty$ limit. Section~\ref{concl} is devoted to some conclusions
and perspectives.

\section{Two ways of obtaining microcanonical averages}
\label{twow}
In the computation of the entropy and of the averages of observables, the problem at hand
is the calculation of integrals of functions of the phase space, restricted to a given
energy shell, and of the dependence of these integrals on the energy. Here we consider two
procedures that are somewhat complementary. They have been recently introduced in the
literature (Refs. ~\cite{rugh1,rugh2} for the finite $N$ approach and
Refs. \cite{ellisa,dauxr} for the large deviation one), and our aim is here 
to make a comparison between them, both as far as general aspects are concerned, 
and for the actual implementation in a specific case.

Rugh's microcanonical formalism~\cite{rugh1,rugh2} is in principle more general, since it is
formulated for general Hamiltonians and for any number of degrees of freedom $N$. It does not allow
to obtain the entropy itself, but it gives access to averages of the observables and to
derivatives of the entropy and of the averages with respect to the energy and to
the parameters of the Hamiltonian. These derivatives are in turn expressed as averages of
suitable observables, which must be determined otherwise, e.g. numerically. On the other hand,
the large deviation method~\cite{ellisa,dauxr} is mostly applicable to systems with long-range interactions.
Expressions are obtained in the thermodynamic limit, although finite $N$
corrections can in principle be derived as series in $1/N$. Its advantage is the
possibility to actually compute the entropy itself, besides its derivatives, and to obtain analytically features of the $N\to\infty$ limit, like phase transitions.

\subsection{Rugh's microcanonical averages}\label{avgen}

In this Subsection we review the basic expressions for the averages of observables in the
microcanonical ensemble, as given in Refs.~\cite{rugh1,rugh2}. We restrict to the case
in which the total energy is the only integral of motion, and we consider a Hamiltonian that 
can depend on a parameter $\lambda$.

In units where the Boltzmann constant is equal to $1$, the microcanonical volume is given by:
\begin{equation}\label{entrm}
\Omega(E,\lambda)={\rm e}^{S(E;\lambda)} = \int {\rm d}\Gamma \, \delta\left(E-H(\Gamma;\lambda)\right) \, ,
\end{equation}
where $\Gamma$ denotes the phase space of the system, and $\lambda$ the parameter the
Hamiltonian depends on. The average of an observable $A(\Gamma)$ is:
\begin{equation}\label{aveg}
\langle A(\lambda) \rangle_{E,\lambda} =\frac{\int{\rm d}\Gamma \, \delta\left(E-H(\Gamma;\lambda)
\right)A(\Gamma;\lambda)}{\int{\rm d}\Gamma \, \delta\left(E-H(\Gamma;\lambda)\right)}~.
\end{equation}
It is possible to express in turn the derivatives with respect to $E$ and $\lambda$ of
$\langle A(\lambda) \rangle_{E,\lambda}$ and of $S(E;\lambda)$ (and thus also the temperature) as averages.
It is useful to introduce the notation
\begin{equation}\label{notat}
\mu(A(\lambda);E,\lambda) = \int{\rm d}\Gamma \, \delta\left(E-H(\Gamma;\lambda)\right)A(\Gamma;\lambda).
\end{equation}
The derivatives of the function $\mu$ with respect to $E$ and $\lambda$ can be expressed using any
vector $\mathbf{Y}$ in $\Gamma$ space such that $\sum_i Y_i \frac{\partial
H}{\partial x_i}\equiv \mathbf{Y} \cdot \mbox{\boldmath $\nabla$} H =1$ (where $x_i$ are the
phase space variables and the gradient operator $\nabla$ acts on $\Gamma$). Using the property
\begin{eqnarray}\label{property}
\sum_i Y_i \frac{\partial}{\partial x_i}\delta (E-H(\Gamma;\lambda))&=&-\delta'(E-H(\Gamma;\lambda))
\sum_i Y_i \frac{\partial H(\Gamma;\lambda)}{\partial x_i}\nonumber \\&=&-\delta'(E-H(\Gamma;\lambda)),
\end{eqnarray}
one gets
\begin{eqnarray}\label{derivmuE}
\frac{\partial}{\partial E}\mu(A(\lambda);E,\lambda)&=&\int {\rm d}\Gamma \delta'(E-H(\Gamma;\lambda))A(\Gamma,\lambda)
\nonumber\\&=&-\int {\rm d}\Gamma \left[\sum_iY_i\frac{\partial}{\partial x_i}\delta (E-H(\Gamma;\lambda))\right]
A(\Gamma;\lambda)\nonumber\\&=&\int {\rm d}\Gamma \delta (E-H(\Gamma;\lambda))
\left[\sum_i\frac{\partial}{\partial x_i}Y_iA(\Gamma;\lambda)\right]\nonumber\\&=&
\mu\left(\mbox{\boldmath $\nabla$} \cdot (A\mathbf{Y});E,\lambda\right),
\end{eqnarray}
where we have integrated by parts. Moreover,
\begin{eqnarray}\label{derivmulambda}
\frac{\partial}{\partial \lambda}\mu(A(\lambda);E,\lambda)&=&
\int {\rm d}\Gamma \left[-\delta'(E-H(\Gamma;\lambda))\frac{\partial H(\Gamma;\lambda)}{\partial \lambda}
\right. \nonumber \\ && \left. +
\delta (E-H(\Gamma;\lambda))\frac{\partial A(\Gamma;\lambda)}{\partial \lambda}\right]\nonumber\\
=\int {\rm d} \Gamma \delta (E-H(\Gamma;\lambda))&&\! \! \! \! \! \! \!
\left[-\sum_i\frac{\partial}{\partial x_i}Y_i
\frac{\partial H}{\partial\lambda}A(\Gamma;\lambda) + 
\frac{\partial A(\Gamma;\lambda)}{\partial \lambda}\right]\nonumber\\
&=&-\mu\left(\mbox{\boldmath $\nabla$} \cdot
(\frac{\partial H}{\partial \lambda}A\mathbf{Y});E,\lambda\right)+\mu\left(\frac{\partial A}{\partial
\lambda};E,\lambda\right)~.
\end{eqnarray}
The entropy $S(E;\lambda)$ and the average $\langle A(\lambda) \rangle_{E,\lambda}$ are expressed,
through the use of~(\ref{notat}), by
\begin{eqnarray}\label{danot}
S(E;\lambda)&=&\ln \mu(1;E,\lambda) \nonumber \\
\langle A(\lambda) \rangle_{E,\lambda}&=&\frac{\mu(A(\lambda);E,\lambda)}{\mu(1;E,\lambda)}.
\end{eqnarray}
From Eqs.~(\ref{derivmuE}), (\ref{derivmulambda}) and~(\ref{danot}) it follows that
\begin{equation}\label{davega}
\frac{\partial}{\partial E}S(E,\lambda)\equiv \frac{1}{T(E,\lambda)} = \langle \mbox{\boldmath $\nabla$}
\cdot \mathbf{Y}\rangle_{E,\lambda}
\end{equation}
\begin{equation}\label{davegb}
\frac{\partial}{\partial \lambda}S(E,\lambda)= -\langle \mbox{\boldmath $\nabla$}
\cdot \left(\frac{\partial H}{\partial \lambda} \mathbf{Y}\right) \rangle_{E,\lambda}
\end{equation}
\begin{equation}\label{davegc}
\frac{\partial}{\partial E}\langle A(\lambda) \rangle_{E,\lambda}=\langle \mbox{\boldmath $\nabla$} \cdot
(A\mathbf{Y}) \rangle_{E,\lambda}-\frac{1}{T(E,\lambda)}\langle A \rangle_{E,\lambda}
\end{equation}
\begin{eqnarray}\label{davegd}
\frac{\partial}{\partial \lambda}\langle A(\lambda ) \rangle_{E,\lambda}&=&-\langle \mbox{\boldmath $\nabla$}
\cdot (\frac{\partial H}{\partial \lambda}A\mathbf{Y}) \rangle_{E,\lambda}
+\langle \mbox{\boldmath $\nabla$} \cdot (\frac{\partial H}{\partial \lambda}\mathbf{Y})
\rangle_{E,\lambda}\langle A \rangle_{E,\lambda} \nonumber \\
&+& \langle \frac{\partial A(\lambda )}{\partial \lambda}
\rangle_{E,\lambda}
\end{eqnarray}

In the simulations, ensemble averages are computed through time averages. This is allowed when
the ergodic hypothesis is satisfied. 
However, it can be argued that, from the practical point of view, this is also allowed in the
presence of metastable states. In fact, in such cases the phase-space trajectory is confined for a long time
(longer than the duration of a simulation) in a subset of the constant energy surface. 
Time averages computed along this trajectory can be interpreted as representative of 
ensemble averages in which another integral of motion, beyond the energy, is present, i.e. 
the characteristic function of the subset where the motion is limited.

\subsection{Entropy from large deviations}
\label{largedev}

Being our system of the mean-field type (see next Section), its microcanonical entropy, given by Eq.~(\ref{entrm}), can be also obtained using large deviation techniques~\cite{ellisa}. 
These techniques are more generally suitable for systems with long-range 
interactions~\cite{LesHouches}. 
It is interesting to derive ensemble averages also in this framework. 

Let us just begin with a brief illustration of how entropy is computed
using large deviation techniques. We suppose that the Hamiltonian of the 
system with $N$ degrees of freedom can be written in the following 
mean-field like form:
\begin{equation}\label{hamlarg}
H = N \sum_{k=1}^n g_k(M_k),
\end{equation}
with
\begin{equation}\label{largvar}
M_k = \frac{1}{N} \sum_{i=1}^N f_k(q_i,p_i),
\end{equation}
and where $f_k,g_k$ in~(\ref{hamlarg}) and ~(\ref{largvar})  are smooth functions.
Leaving implicit the dependence on quantities other than the energy, the microcanonical volume, Eq.~(\ref{entrm}), can be transformed as:
\begin{eqnarray}\label{entrlarg}
&{\rm e}^{S(N\epsilon)} &= \int {\rm d}\Gamma \, \delta\left(H(\Gamma)-E\right)
= \int {\rm d}\Gamma \, \delta\left(N \sum_{k=1}^n g_k(M_k(\Gamma)) -N\epsilon\right)= \nonumber \\
&=&\int {\rm d}\Gamma \left[\Pi_{k=1}^n{\rm d}m_k \, \delta(NM_k(\Gamma) -Nm_k)\right]
\delta\left[N\left( \sum_{k=1}^n g_k(m_k) -\epsilon\right)\right]~.
\end{eqnarray}
Using the inverse Laplace transform of the $\delta$ function we get:
\begin{eqnarray}\label{entrlarg1}
{\rm e}^{S(N\epsilon)} &=& \int {\rm d}\Gamma \left[\Pi_{k=1}^n{\rm d}m_k{\rm d}z_k\right] \,
\left(\frac{N}{2\pi {\rm i}}\right)^n \delta\left[N\left( \sum_{k=1}^n g_k(m_k) - \epsilon\right)\right]
\cdot \nonumber \\
&\cdot& \exp \left[\sum_{k=1}^n\left( Nz_kM_k(\Gamma) -Nz_km_k\right)\right]= \nonumber \\
&=& \int \left[\Pi_{k=1}^n{\rm d}m_k{\rm d}z_k\right] \,
\left(\frac{N}{2\pi {\rm i}}\right)^n \delta\left[N\left( \sum_{k=1}^n g_k(m_k) - \epsilon\right)\right]
\cdot \nonumber \\
&\cdot& \exp \left\{-N\left[\left(\sum_{k=1}^nz_km_k\right)-\ln \langle
{\rm e}^{\sum_{k=1}^n z_kf_k}\rangle_1\right]\right\}~,
\end{eqnarray}
where $\langle \cdot \rangle_1$ is the one particle phase-space $(q,p)$ average given by:
\begin{equation}\label{oneaver}
\langle \cdot \rangle_1 = \int {\rm d}q{\rm d}p \, (\cdot)~,
\end{equation}
which is of course defined only for integrable functions. In Eq.~(\ref{entrlarg1}) the integration path
for the variables $z_k$ is a line parallel to the imaginary axis and with a positive real part.
The integrals in $z_k$ are computed, for $N\rightarrow \infty$, using the saddle point method.
The relevant stationary point of the function in square brackets in the exponent in~(\ref{entrlarg1}) must lie on the real $z_k$ axes. Otherwise one would find an unphysical oscillatory behaviour of the integral. The minimum along a line parallel to the imaginary axis is a maximum along the
real axis; therefore we get:
\begin{eqnarray}\label{entrlarg2}
{\rm e}^{S(N\epsilon)} &=& \int \left[\Pi_{k=1}^n{\rm d}m_k\right] \,
\left(\frac{N}{2\pi}\right)^n \delta\left[N\left( \sum_{k=1}^n g_k(m_k) - \epsilon\right)\right]
\cdot \nonumber \\ &\cdot&
\exp \left[-N I\left(m_1,\dots,m_n\right)\right]~,
\end{eqnarray}
where:
\begin{equation}\label{supi}
I\left(m_1,\dots,m_n\right) = \sup_{z_1,\dots,z_k}\left[\left(\sum_{k=1}^nz_km_k\right)-\ln \langle
{\rm e}^{\sum_{K=1}^n z_kf_k}\rangle_1\right]~,
\end{equation}
with the $\sup$ taken for real values of the $z_k$'s. Finally, in the $N\to\infty$
limit we have:
\begin{equation}\label{entrlarg3}
s(\epsilon) \equiv \lim_{N\to\infty} \frac{1}{N}S(N\epsilon) = -I(\epsilon)~,
\end{equation}
where:
\begin{eqnarray}\label{infi}
I(\epsilon)&=& \inf_{m_1,\dots,m_n,\sum g_k(m_k)=\epsilon} I\left(m_1,\dots,m_n\right)= \nonumber
\\ &=& -\sup_{m_1,\dots,m_n,\sum g_k(m_k)=\epsilon}\left[-I\left(m_1,\dots,m_n\right)\right]~.
\end{eqnarray}
If the last $\inf$ is taken only on some of the $m_k$, then we obtain the entropy as a function
of $\epsilon$ and of the remaining $m_k$'s, which are then considered as given quantities.

Later, in Section~\ref{implem}, we will implement both
procedures in our system. Since expressions
computed using large deviation techniques describe the system in the $N\to\infty$ limit, the 
comparison will be performed with formulas obtained with  Rugh's microcanonical 
formalism taken in the same limit.

The important point to be noted is the following. Rugh's formalism is more general, 
since it is valid, in principle, for any system and for any size $N$. It provides expressions for the average of observables and for the derivatives of these averages, that afterwards can be effectively
computed in simulations. However, the microcanonical volume, and the entropy itself, can not be evaluated. On the other hand, large deviation techniques are
more easily applied to systems with long-range interactions, but one can obtain an explicit expression of the entropy, which gives access to the full
knowledge of thermodynamic properties, including phase transitions.
\section{The $\phi^4$ model in an external field}\label{model}

The Hamiltonian of our system is given by:
\begin{equation}\label{ham}
H = K+V = \sum_{i=1}^N \left[\frac{p_i^2}{2}-(1-\theta)\frac{1}{2}q_i^2 + \frac{1}{4}q_i^4 -h q_i
\right] -\frac{\theta}{2N} \sum_{i,j=1}^N q_i q_j~,
\end{equation}
where the $q_i$ variables can be thought as describing the position of the $i$-th 
particle on a line, and the $p_i$'s are the conjugate momenta (the mass is unitary). 
Each particle is subject to both a local potential and to an infinite range one, expressed by the all-to-all coupling; $h$ is the external magnetic field and $\theta$ is a free parameter.
Indeed, it can be shown that all parameters appearing in a potential
energy of this form can be absorbed in $\theta$ by a convenient change of
variables. In the MD simulations we have chosen the
value $\theta=1/2$, which is in the range that ensures the presence of an effective double well potential in the low energy phase. 
The Langevin dynamics with the force coming from the potential energy
of model~(\ref{ham}) has been solved in Ref.~\cite{zwanzig}. A careful study of the different dynamical regimes for the relaxation to equilibrium and a first characterization of the second order phase transition has been performed in the same paper. 
A further study of the canonical ensemble solution of this model for $h=0$ 
has been performed in Ref.~\cite{thierry}, where it has been shown that the system displays, for $\theta>0$, a second order ferromagnetic transition.
The magnetization defined by
\begin{equation}\label{mdef}
M_1 = \frac{1}{N}\sum_{i=1}^N q_i
\end{equation}
vanishes continuously at a critical temperature, which can be determined numerically 
by solving an implicit equation. In Ref.~\cite{thierry} the authors have shown 
that for $\theta=1/2$ the critical temperature is $T_c \simeq 0.264$. For this 
particular value of $\theta$ it can be shown that, curiously, the corresponding 
critical energy density $\epsilon_c$ is equal to $T_c/2$, thus $\epsilon_c \simeq 0.132$. 

Hamiltonian~(\ref{ham}) can be cast in the form~(\ref{hamlarg}):
\begin{equation}\label{haml1}
H=N\left[ \frac{1}{2}P-\left( hM_1+\frac{\theta}{2}M_1^2\right)-\frac{1}{2}(1-\theta)M_2
+\frac{1}{4}M_4 \right]~,
\end{equation}
using the $n=4$ mean-fields (see Eq.~(\ref{largvar}))
\begin{equation}\label{coarsdef}
P=\frac{1}{N}\sum_{i=1}^N p_i^2 \quad \quad \quad \quad \quad M_s=\frac{1}{N}\sum_{i=1}^N q_i^s,
\end{equation}
with $s=1,2,4$.
\section{Implementation of the general expressions}\label{implem}

In this Section we will implement the general expressions introduced in Section~\ref{twow},
discussing the solution of the mean-field $\phi^4$ model in the microcanonical ensemble.

\subsection{Finite $N$ microcanonical averages}\label{averages}
Following Ref.~\cite{rugh1}, we take for $\mathbf{Y}$ the vector given by:
\begin{equation}\label{yfree}
\mathbf{Y} = \frac{1}{\sum_{i=1}^N p_i^2}\left(p_1,\dots,p_N,0,\dots,0\right) \equiv
\frac{1}{2K}\left(p_1,\dots,p_N,0,\dots,0\right),
\end{equation}
with non vanishing components only in correspondence of the momenta. The external magnetic
field $h$ plays the role of the parameter $\lambda$. We are interested in the expressions
for the temperature, the specific heat, the average magnetization and the magnetic
susceptibility. Computing $\nabla \cdot \mathbf{Y}$, the temperature follows from
Eq.~(\ref{davega}):
\begin{equation}\label{tfree}
\frac{1}{T(E,h)} = \langle \frac{N-2}{2K}\rangle_{E,h}.
\end{equation}
From this, and using~(\ref{davegb}), the inverse specific heat is given by:
\begin{eqnarray}\label{shfree}
\frac{1}{C(E,h)}&=& \frac{\partial}{\partial E}T(E,h) = -T^2(E,h)\frac{\partial}{\partial E}
\frac{1}{T(E,h)} = \\
&=& 1 - T^2(E,h)\langle \frac{(N-2)(N-4)}{4K^2}\rangle_{E,h}
= 1 - \frac{(N-4)}{(N-2)}\frac{\langle 1/K^2 \rangle_{E,h}}{\langle 1/K \rangle_{E,h}^2}~. \nonumber
\end{eqnarray}
We point out that $C$ is the extensive specific heat and thus
Eq.~(\ref{shfree}) is expected to scale as $N^{-1}$.
The average magnetization $\langle M_1 \rangle_{E,h}$, with $M_1$ given by~(\ref{mdef}), is
denoted by $m(E,h)$. Then the magnetic susceptibility at constant energy is computed
through~(\ref{davegc}), noting that $\partial H/\partial h = -NM_1$, and it is expressed by:
\begin{equation}\label{suscfree}
\chi (E,h)=\frac{\partial}{\partial h} m(E,h) = N(N-2)\left[ \langle \frac{M_1^2}{2K}
\rangle_{E,h}-\langle \frac{M_1}{2K}\rangle_{E,h}m(E,h)\right]~.
\end{equation}

\subsection{Microcanonical solution of the $\phi^4$ model}
\label{solution}

In this Subsection we will present the solution of the mean-field $\phi^4$ model using large
deviations techniques. In particular, we will derive explicit expressions for the microcanonical 
entropy and its derivatives in the $N \to \infty$ limit. 
We will apply the formalism introduced in Subsection~\ref{largedev}.

The one particle average in the rightmost side of Eq.~(\ref{entrlarg1}) takes the form
\begin{eqnarray}\label{psidef}
\psi(z_p,z_1,z_2,z_4)&=&\int {\rm d}q{\rm d}p \, \exp \left[ z_p p^2 + z_1 q + z_2 q^2 +  z_4 q^4
\right] = \nonumber \\
&=&\sqrt{\frac{\pi}{-z_p}}\int {\rm d}q \, \exp \left[ z_1 q + z_2 q^2 + z_4 q^4 \right]~,
\end{eqnarray}
where convergence requires that $z_p$ and $z_4$ be negative. 
The function $\psi(z_p,z_1,z_2,z_4)$ is the so-called {\it generating function} in
large deviation theory.

Therefore, the extremal problem defined by~(\ref{supi}) is:
\begin{eqnarray}\label{supidef}
\! \! \! \! \! \! \! \! \!
I(u,Q_1,Q_2,Q_4) &=& \sup_{z_p,z_1,z_2,z_4}\left[ z_pu+z_1Q_1+z_2Q_2+z_4Q_4 - \ln
\psi(z_p,z_1,z_2,z_4)\right]~, \nonumber \\ &&\phantom{1}
\end{eqnarray}
where the symbols $u$ and $Q_1$, $Q_2$ and $Q_4$ stand for the $m_k$ in~(\ref{entrlarg1}). The
variable $z_p$ separates from the others and, after straightforward 
calculation, one obtains:
\begin{eqnarray}\label{supcalc}
&&I(u,Q_1,Q_2,Q_4)= -\frac{1}{2}\ln (2\pi {\rm e})-\frac{1}{2}\ln u +\overline{z}_1(Q_1,Q_2,Q_4)Q_1
+ \nonumber \\ &+& \overline{z}_2(Q_1,Q_2,Q_4)Q_2 + \overline{z}_4(Q_1,Q_2,Q_4)Q_4
-\ln G(\overline{z}_1,\overline{z}_2,\overline{z}_4)~,
\end{eqnarray}
where the function $G$ is given by:
\begin{equation}\label{gdef}
G(z_1,z_2,z_4)=\int {\rm d}q \, \exp \left[ z_1 q + z_2 q^2 + z_4 q^4 \right]~,
\end{equation}
and the functions $\overline{z}_s(Q_1,Q_2,Q_4)$ are the solution of:
\begin{equation}\label{extremq}
Q_s= \frac{1}{G}\frac{\partial}{\partial z_s}G(z_1,z_2,z_4) \equiv \frac{G_s}{G}(z_1,z_2,z_4)
\end{equation}
for $s=1,2,4$. Here $G_s$ is expressed by:
\begin{equation}\label{gsdef}
G_s(z_1,z_2,z_4)=\int {\rm d}q \, q^s\exp \left[ z_1 q + z_2 q^2 + z_4 q^4 \right]~.
\end{equation}
In the actual computation one has to check that the stationary points found from~(\ref{extremq})
correspond to the proper extrema. Finally, using~(\ref{entrlarg3}) and~(\ref{infi}), the entropy
is:
\begin{eqnarray}\label{unconentr}
s(\epsilon,h)&=& \sup_{Q_1,Q_2,Q_4} \left\{ \frac{1}{2}\ln \left[ 2\epsilon + (2hQ_1+\theta Q_1^2)
+(1-\theta)Q_2 -\frac{1}{2}Q_4 \right] + \right. \nonumber \\
&& \left. \phantom{\frac{1}{2}} + \ln G(\overline{z}_1,\overline{z}_2,\overline{z}_4)-\overline{z}_1Q_1
-\overline{z}_2Q_2-\overline{z}_4Q_4 \right\}~.
\end{eqnarray}
Here we have used the equation of the total energy to eliminate the variable $u$.
Using Eq.~(\ref{extremq}), that defines the functions $\overline{z}_s(Q_1,Q_2,Q_4)$,
the stationary points that solve the variational problem~(\ref{unconentr}) are:
\begin{eqnarray}\label{statz}
\frac{\theta Q_1+h}{u} &=& \overline{z}_1 \nonumber \\
\frac{1-\theta}{2u} &=& \overline{z}_2 \nonumber \\
-\frac{1}{4u} &=& \overline{z}_4~,
\end{eqnarray}
where, for brevity, the expression in square brackets in~(\ref{unconentr}) has again been denoted
as $u$.
The solution for $Q_s$ (and $u$) can be denoted with $\overline{Q}_s$ (and $\overline{u}$).
It can be seen that, as expected, $\overline{z}_4$ is a negative quantity, since the kinetic
energy per particle $\frac{1}{2}\overline{u}$ is positive. Also, one has that
$\overline{z}_2=-2(1-\theta)\overline{z}_4$.

The first derivatives of the entropy can now be explicitly computed, using well
known thermodynamic relations. Taking into account the stationary point in 
Eqs.~(\ref{extremq}) and~(\ref{statz}), one gets: 
\begin{eqnarray}\label{equncon}
\frac{\partial s(\epsilon,h)}{\partial \epsilon}&\equiv& \frac{1}{T}=\frac{1}{\overline{u}}
\nonumber \\
\frac{\partial s(\epsilon,h)}{\partial h}&\equiv& \frac{m}{T}=\frac{\overline{Q}_1}{\overline{u}}~.
\end{eqnarray}

The first equation is exactly the same as Eq.~(\ref{tfree}), which is obtained
using Rugh's formalism, once it is taken in the thermodynamic
limit. The second of the Eqs.~(\ref{equncon}) is practically a tautology, since it can be shown that the
$\overline{Q}_s$ are simply the averages of the quantities $M_s$ defined in Eq.~(\ref{coarsdef}). We
also note that the functions $\overline{z}_2$ and $\overline{z}_4$ can be expressed as
$\overline{z}_2=(1-\theta)\frac{1}{2T}$ and $\overline{z}_4=-\frac{1}{4T}$.

Before considering the second derivatives, we show how to obtain the critical point. 
We consider the system with $h=0$. Then, Eqs.~(\ref{extremq}) and~(\ref{statz})
have always the solution $\overline{z}_1=\overline{Q}_1=0$. However, for an energy density $\epsilon$
smaller than a critical energy $\epsilon_c$, the relevant solution is one with $\overline{Q}_1>0$,
although we do not prove it here. We only show how to obtain the value of $\epsilon_c$ and the
corresponding critical temperature $T_c$.

From Eq.~(\ref{statz}) one gets the relations $\overline{z}_4=-\overline{z}_2/(2-2\theta)$
(already mentioned) and $\overline{z}_1=2\theta\overline{Q}_1\overline{z}_2/(1-\theta)$ (for
$h=0$). One can insert these relations in Eq.~(\ref{extremq}). Near the critical point,
$\overline{Q}_1$ is small; we can therefore write Eq.~(\ref{extremq}) for $s=1$, performing
a power series expansion in $\overline{Q}_1$ up to first order. Calling $\overline{z}_{2c}$ the
critical value of $\overline{z}_2$, we easily obtain the following equation for
$\overline{z}_{2c}$:
\begin{equation}\label{z2cdef}
\frac{\frac{2\theta}{1-\theta}\overline{z}_{2c}\int {\rm d}q \, q^2 \exp \left\{
\overline{z}_{2c}\left[q^2-\frac{1}{2(1-\theta)}q^4\right]\right\}}
{\int {\rm d}q \, \exp \left\{\overline{z}_{2c}\left[q^2-\frac{1}{2(1-\theta)}q^4\right]\right\}}
=1~,
\end{equation}
which can be easily solved numerically. This same equation also tells that:
\begin{equation}\label{q2cdef}
\overline{Q}_{2c}=\frac{1-\theta}{2\theta \overline{z}_{2c}}~.
\end{equation}
It can be also easily shown that:
\begin{equation}\label{q4cdef}
\overline{Q}_{4c}=\overline{Q}_{2c}~.
\end{equation}
From the second of Eqs.~(\ref{statz}) and from the Hamiltonian, the critical temperature 
and the critical energy are found to be :
\begin{eqnarray}\label{crittemp}
T_c&=&\overline{u}_c=\frac{1-\theta}{2\overline{z}_{2c}} \nonumber \\
\epsilon_c&=&\frac{(1-\theta)(2\theta-\frac{1}{2})}{4\theta \overline{z}_{2c}}~. \nonumber \\
\end{eqnarray}

Let us now consider the second derivatives of the entropy and, to be definite, 
let us consider the specific heat computed from the entropy ~(\ref{unconentr}). We do not show
here the cumbersome expression obtained differentiating twice the entropy~(\ref{unconentr}) with respect
to $\epsilon$. However, it can be easily guessed that several terms containing the coordinates,
and not only the momenta, will appear. This is in contrast with Eq.~(\ref{shfree})
derived from Rugh's formalism, where only
the kinetic energy appears. Of course the results must be equivalent, but the connection between
kinetic and potential energy in the microcanonical ensemble (absent in the canonical ensemble)
hinders the equivalence. However, an equation analogous to~(\ref{shfree}), when this is taken in
the thermodynamic limit, can be still obtained if one performs the derivatives before performing
the integrations leading to Eq.~(\ref{unconentr}). Let us focus on the $\delta$ function in the
rightmost hand side of~(\ref{entrlarg1}). For our system its argument is:
\begin{equation}\label{argdelt}
N\left[ \frac{1}{2}u-\frac{\theta}{2}Q_1^2-\frac{1}{2}(1-\theta)Q_2
+\frac{1}{4}Q_4-hQ_1-\epsilon\right]~.
\end{equation}
Thus, the derivative with respect to $\epsilon$ of the $\delta$ function is equal to
$-2$ times the derivative with respect to $u$. Integrating by parts in $u$ we then find that:
\begin{equation}\label{invsh1}
\frac{\partial^2 S(N\epsilon)}{\partial (N\epsilon)^2}=
4\left[ \langle z_p^2 \rangle - \langle z_p \rangle^2 \right]~,
\end{equation}
with the microcanonical average $\langle \cdot \rangle$ computed through the microcanonical
formula~(\ref{entrlarg1}). The saddle point evaluation of~(\ref{invsh1})
has to be performed up to order $O(N^{-1})$, since
the terms of $O(1)$ cancel out. The calculation, after several steps, leads to:
\begin{equation}\label{invsh2}
\frac{\partial^2 S(N\epsilon)}{\partial (N\epsilon)^2}=
-\frac{2}{N\langle u \rangle^2}+\frac{\langle (\delta u)^2 \rangle}{\langle u \rangle^4}~,
\end{equation}
where $u$, as before, is twice the kinetic energy, and $\langle (\delta u)^2 \rangle$ denotes
the variance of $u$. Therefore the inverse specific heat is:
\begin{equation}\label{invsh3}
\frac{1}{C}=-T^2\frac{\partial^2 S(N\epsilon)}{\partial (N\epsilon)^2}=
\frac{2}{N}-\frac{\langle (\delta u)^2 \rangle}{\langle u \rangle^2}~.
\end{equation}
In~(\ref{invsh2}) and~(\ref{invsh3}) the two terms are both of order $N^{-1}$, in spite
of their appearence. It can be easily shown~\cite{rugh1} that the $N\to\infty$ limit 
of~(\ref{shfree}) leads to exactly the same expression.

The same kind of procedure can be followed for the magnetic susceptibility. Now we can use 
the fact that the derivative with respect to $h$ of the $\delta$ function in the rightmost 
hand side of~(\ref{entrlarg1}) is equal to $-2Q_1$ times the derivative with respect 
to $u$. Integrating again by parts we get:
\begin{equation}\label{invsh4}
\chi = -2N\left[ \langle Q_1^2z_p \rangle -m \langle Q_1z_p \rangle \right]~.
\end{equation}
The leading order in $N$ of this expression gives:
\begin{equation}\label{invsh5}
\chi = \frac{N}{\langle u \rangle}\left[ \langle Q_1^2 \rangle -m\langle Q_1 \rangle \right]~,
\end{equation}
which is the same as the thermodynamic limit of~(\ref{suscfree}), reminding that $Q_1$
corresponds to the dynamical variable $M_1$ given in Eq.~(\ref{mdef}).

\section{Molecular dynamics simulations}\label{results}

In this Section we show some results of MD simulations of our system. In
particular, we present numerical data on the microcanonical temperature, the specific heat,
the magnetization and the magnetic susceptibility.

The simulations have been performed with $N=25,50,100,200$ particles, and the value
$\theta=1/2$ has been selected. We have considered here only energies above the critical energy
$\epsilon_c \simeq 0.132$ (see Section~\ref{model}), since in the simulations presented 
in this work we only want to show that the microcanonical formalism
can be easily applied, in practical numerical computations. Below the critical energy there can
be metastable states~\cite{thierry}, that we do not consider here. In a work in preparation,
devoted to nonequivalence of statistical ensembles, we will also present the results 
of numerical simulations performed below the critical energy~\cite{CRT}.

We have found that the convergence of the averages can be significantly improved if, 
during the simulation, the velocities of the particles are reshuffled at given 
time intervals. This is particularly true for the quantities related to the 
second derivatives of the entropy, i.e. the specific heat and the magnetic susceptibility. 
The faster convergence is presumably due to a more efficient spanning of the phase space 
of the system.

\begin{figure}[htbp]
  \centering
\includegraphics[width=14.0cm,height=17.0cm]{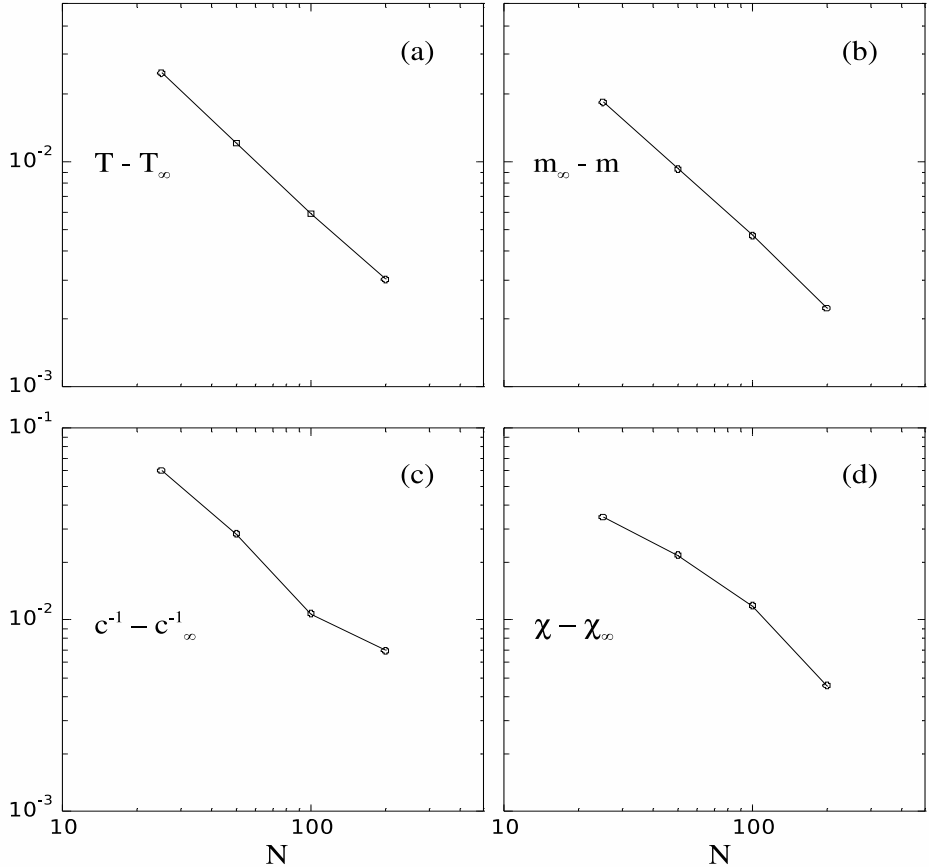}
  \caption{Microcanonical time average of temperature $T$ (a), magnetization $m$(b),
  inverse specific heat $1/c$ (c), magnetic susceptibility $\chi$(d). In all figures we
  show the difference of the time average at finite $N$ with respect to the 
  $N\to\infty$ limit value (which is analytically computed using large deviation 
  techniques and is indicated by the $\infty$ subscript) vs. the number $N$ 
  of particles. The energy density is $\epsilon = 0.25$}
  \label{fig1}
\end{figure}

In Fig.~\ref{fig1} we show the results for the energy density $\epsilon = 0.25$. 
The simulation time is $2\cdot 10^5$. The figure contains four plots in log-log scale, in which
we show, as a function of the number of particles, the difference between the average value
of four observables obtained in the finite $N$ simulations and their corresponding 
$N\to\infty$ limit
computed through the large deviation method (indicated by the subscript $\infty$). 
The four observables are: the microcanonical
temperature $T$~(\ref{tfree}), the magnetization $m = \langle M_1 \rangle_{E,h}$~(\ref{mdef}), 
the inverse specific heat $1/c$~(\ref{shfree}) and the magnetic susceptibility 
$\chi$~(\ref{suscfree}). Actually, for the
magnetization, the plotted quantity is the $N\to\infty$ limit value minus the finite $N$ value,
which is smaller. Furthermore, for the inverse specific heat we have considered the intensive quantities
that are obtained multiplying~(\ref{shfree}) and~(\ref{invsh3}) by $N$ (and then we have used the
lower case for the corresponding symbol).

Assuming a power law decay with $N$ of the simulated averages towards the $N\to\infty$ limit
value, the exponent of this law can be estimated from a fit with a straight line. 
In the plots the lines are interpolations between points, as a guide to the eye. We do not
show the fitting lines that can be easily obtained. We limit to point out that the slope of these
lines is in all cases close to $-1$. This behavior should be expected on general grounds; in fact,
the finite $N$ corrections could be obtained by the successive terms in the saddle point evaluation
related to the large deviation computation, and the first correction is expected in general to scale 
like $N^{-1}$.

\section{Conclusions and perspectives}\label{concl}
In this paper we have considered in parallel two methods for the evaluation of 
microcanonical averages of macroscopic observables and we have applied them to 
the mean-field $\phi^4$ model in an external magnetic field.

We have implemented in the model the general expressions that, 
through the microcanonical formalism  introduced by Rugh~\cite{rugh1,rugh2}, 
allow us to compute an entire set of macroscopic  quantities (temperature, magnetization, 
specific heat, magnetic susceptibility, etc.), performing time averages of suitable 
mechanical observables at finite $N$. 
In particular, these macroscopic variables are obtained taking derivatives 
of the entropy with respect to the energy density and to other parameters 
of the Hamiltonian, like the external field (see e.g. formulas~(\ref{equncon})). 

Then, we have considered large deviation techniques, whose applicability relies
on the determination of a set of macroscopic variables on which the energy
depends in a simple way. This is generally possible for systems with long-range 
interactions, although the method is viable also for other cases~\cite{ellisa}. 
Our model, being of the mean-field type, can be solved exactly in
the microcanonical ensemble, using large deviation techniques. This leads 
to the analytic determination of the same macroscopic quantities in the 
thermodynamic limit.

The purpose was here to derive
the actual expressions of macroscopic observables that stem from the application 
of the two different methods mentioned above to a model that has already been 
a subject of study for  its interesting dynamical properties~\cite{zwanzig,thierry}. 
Although these two approaches lead to expressions which 
show no evident similarity, especially for specific heat and magnetic 
susceptibility, the time averages computed with Rugh's microcanonical formalism
converge to the thermodynamic limit value computed with large deviations.
Corrections are of order $N^{-1}$. 

We find that the specific heat and the magnetic susceptibility converge extremely slowly 
in time, unless one devises methods to ``help'' the system explore more efficiently phase space. 
The method we use consists in periodically reshuffling particle velocities.
These observables, which correspond to higher order derivatives of the entropy, 
have extremely complex expressions,  and this could be at the origin of the observation 
of such slow convergence. However, we believe that this slow relaxation 
is rather a distinctive signature of the long-range (mean-field) nature of the model. 
Indeed, slow time relaxations are quite common in such systems and have been already observed 
in other models~\cite{LesHouches,thierry,Yama,Kaneko}, being associated to the presence of the
so-called {\it quasi-stationary states}.   

In a future paper~\cite{CRT} we will concentrate on the properties of this model
in the low energy/temperature phase. We will show the presence of ensemble inequivalence 
features and of a negative susceptibility in the microcanonical ensemble.

While finishing this paper we became aware of a very recent study of the $\phi^4$ model 
without external field which uses large deviation techniques~\cite{kast}. 
However, this work focuses on different aspects, being devoted to a detailed study 
of the non-analiticity properties of the entropy function.

\begin{ack}
We thank the Laboratoire de Physique of ENS-Lyon for hospitality during the writing
of a first draft of this paper. S.R. acknowledges financial support of the PRIN03 project 
{\it Order and chaos in nonlinear extended systems} and of CNRS and ENS-Lyon.
\end{ack}


\end{document}